# A Detailed Study of Dwarf Galaxies in the Core of the Coma Cluster

Nagamani Poloji[1], Priya Hasan[2,3,*] and S.N.Hasan[4]

[1]Department of Astronomy, Osmania University, Hyderabad, India
[2]Department of Physics, Maulana Azad National Urdu University, Hyderabad, 500032, India
[3]The Inter-University Centre for Astronomy and Astrophysics, Post Bag 4, Ganeshkhind, Pune, Maharashtra 411007, India
[4] Department of Mathematics, Maulana Azad National Urdu University, Hyderabad, 500032, India

*Corresponding author. E-mail: priya.hasan@gmail.com

**Abstract.** In an earlier paper, we determined the morphological types of galaxies in the Coma Cluster using data from the HST/ACS Coma Cluster treasury survey. We found that of the 132 members, 51 are non dwarfs and 81 are dwarfs. We define dwarfs to have a absolute luminosity $M_{F814W} \geq -18.5$ as in Marinova *et al.* (2012). In this paper, we determine the morphological types of these dwarf galaxies and make a detailed study of their properties. Using GALFIT, we determine the structural properties of our sample and with spectroscopic redshifts, we determined memberships and distances to identify dwarfs. A visual examination of the residual images reveals that our sample of 78 dwarf galaxies comprises of: dwarf lenticular (*dS0*) 22%, dwarf Elliptical (*dE*) 69%, dwarf spirals (*dSp*) 4%, dwarf ring (*dring*) 1%, dwarf barred spirals (*dSBp*) 3% and dwarf irregular (*dIrr*) 1% galaxies. We find that the bulge-disk decomposition (Sérsic + exponential) fits are good only for the *dS0* galaxies. The remainder of the sample gives good fits only for single Sérsic fits. The Colour Magnitude Relation (CMR) shows that the *dEs* are redder and fainter than the rest of the sample (except one *dIrr* galaxy). The Kormendy relation reveals that *dE* galaxies have lower surface brightness than the rest of the sample. Our research leads us to the conclusion that dwarf galaxies appear to have a different formation and evolution process than non-dwarf galaxies.

## 1. Introduction

In the standard scenario, the gravitational collapse of primordial density fluctuations produce dwarf galaxies (Dekel & Silk, 1986). After the first stars form, a galaxy's structure can change, regulating ongoing star formation through a variety of energy feedback mechanisms to the interstellar medium. The Colour Magnitude Relation (CMR), is a good indicator of the mean age of galaxies (Hamraz *et al.*, 2019). This implies that in general, blue galaxies are younger, while red galaxies are older. The other important relations we shall study are the relations between surface brightness and effective radius (Kormendy relation (Kormendy, 1977, 1985)) and the relation between surface brightness and absolute magnitude which is used to study the formation and evolution of early type galaxies.

The majority of galaxies in our universe are dwarfs and therefore are key to understanding the structure and evolution of galaxies, especially in clusters. In our earlier papers (Poloji *et al.*, 2022; Poloji *et al.*, 2022, 2023), we studied the galaxies in Coma Cluster, focusing on the non-dwarf galaxies. However, the majority of galaxies in the Coma Cluster are dwarfs (Marinova *et al.*, 2012; Poloji *et al.*, 2022) and especially the Coma cluster core, is dominated by dE galaxies Secker *et al.* (1997) and this paper is devoted to their study.

Dwarf galaxies, in general, have low surface brightness and in this paper we define dwarfs to have a luminosity $M_{F814W} \geq -18.5$ as in Marinova *et al.* (2012). Dwarf galaxies exhibit varied morphologies, similar to their non-dwarf counterparts. The most frequent categories found are dwarf ellipticals (*dE*) and dwarf irregular (*dIrr*) galaxies. In general, due to the low luminosity of these galaxies, it is often difficult to assign morphological types.

At a redshift of 0.023 (100 Mpc), the Coma Cluster is an extremely rich and dense cluster (Hammer *et al.*, 2010). It is also ideal for research on the luminosity, environment, and morphological classification of dwarf galaxies since it is abundant with dwarf galaxies. Numerous morphological investigations of dwarf galaxies in the Coma Cluster have been conducted (Andreon, 1996; Thompson & Gregory, 1993; Komiyama *et al.*, 2002; Gutiérrez *et al.*, 2004; Michard, R. & Andreon, S., 2008; Kourkchi *et al.*, 2012; Marinova *et al.*, 2012; Hasan *et al.*, 2019). Michard, R. & Andreon, S. (2008) have used data from the GMP 1983 catalogue with the CFH12K Camera and studied galaxy properties using isophotal contours for a sample of 43 dwarfs within a magnitude limit of $M_B = -15$. (Hamraz *et al.*, 2019) has obtained the CMR for Coma, Virgo and Fornax, however, without considering morphological types.

In this paper, we make a detailed study of the sample of dwarf galaxies in the core of the Coma Cluster using visual inspection of the science image, model and residual images of GALFIT from single Sérsic profiles as well as bulge-disk (Sérsic + exponential) decomposition, where possible. The residual image is very crucial in identifying spiral arms, bars and other features. We study, in detail, the quantified morphological parameters like Sérsic index $n$, colour, surface brightness, effective radius ($r_e$) and their correlations. We have earlier studied the CMR–Morphology relationship for non-dwarfs and compared them with the dwarfs of the Coma Cluster.

This paper is structured as follows: The introduction is presented in Section 1. The HST-ACS data and membership determination is described in Section 2. Section 3 discusses the Two-dimensional decomposition and the GALFIT results used to determine the structural characteristics of galaxies. Section 4 presents the analysis and discussion based on the CMR, the Colour Distance Relation and the Kormendy relation. Section 5 presents the Conclusions of our paper.

## 2. Data and membership

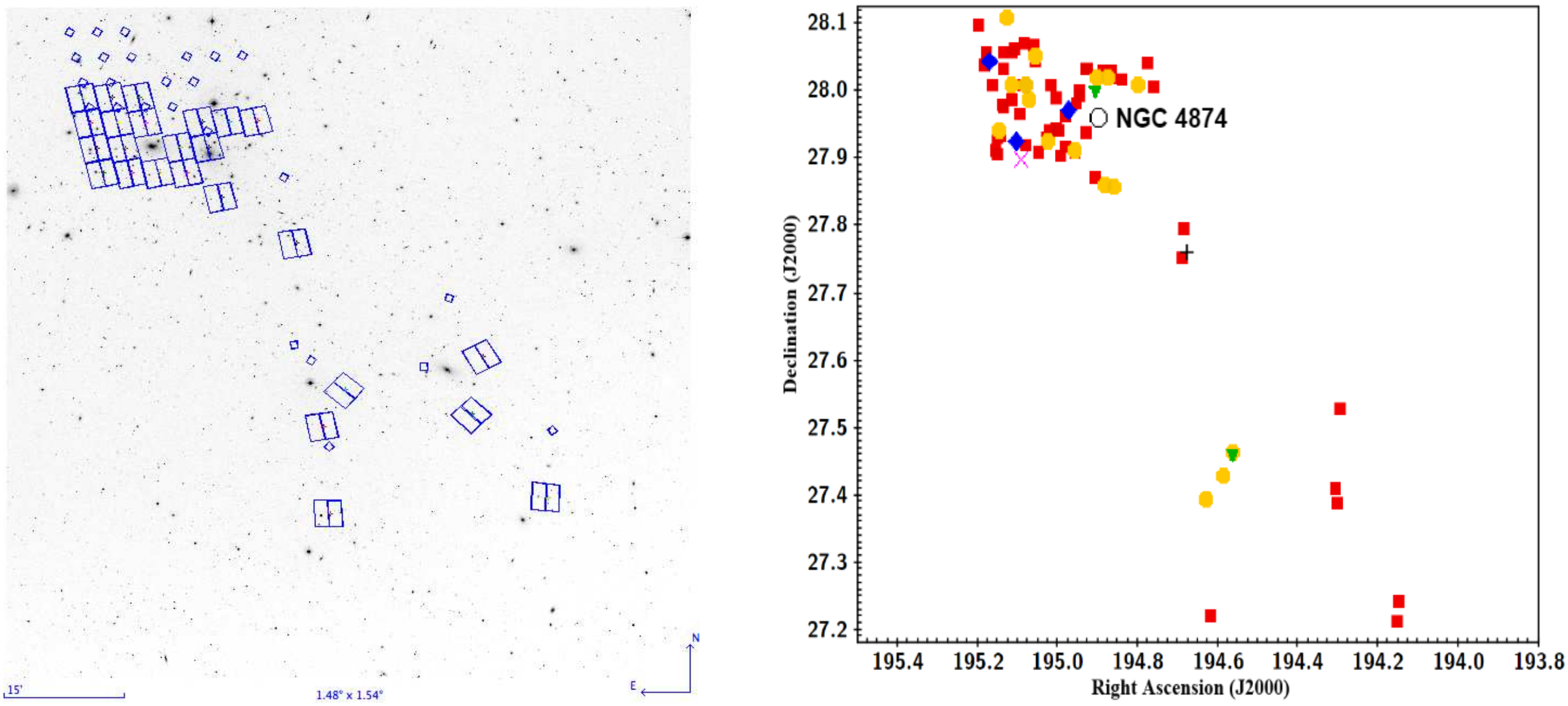

**Figure 1**. The HST-ACS Survey of Coma Cluster with the distribution of dwarf morphological types. The figure on the left shows the locations of the 25 tiles observed. The figure on the right shows the distribution of the morphologies of the sample of dwarf galaxies described in this paper. The red filled squares represents the *dE* galaxies, yellow filled circles are *dS0*, pink cross mark is *dring*, blue filed rhombus are *dSp*, green filled triangles are *dSBp*, black plus mark is *dIrr* and black open circle represents the cD galaxy (NGC 4874).

The HST-ACS[1] Coma Cluster Treasury Survey data release 2.1 provided us with publicly accessible data (Carter *et al.*, 2008). The survey was initially intended to cover a section of the cluster with a 740 arcmin$^2$ area in regions with different densities of galaxies and intergalactic space. Figure 1 shows the location of the 25 tiles observed on the left. The figure on the right shows the distribution of the morphologies of the sample of dwarf galaxies described in this paper. With exposure times of 1400 and 2560 seconds, it has data in the filters $F814W$ and $F475W$. The survey was only 28% complete due to an ACS failure on 27th January 2007. A total of 25 ACS tiles were covered, of which 19 tiles were within 0.5 Mpc and the remaining 6 tiles were from 0.9 to 1.75 Mpc from the centre. The observations span 25 fields, of which 19 are within 0.5 Mpc of the cluster's centre and the remaining 6 are near its south-west extension.

The science images and the SourceExtractor catalogues (SExtractor version 2.5; Bertin & Arnouts (1996)) for the 25 fields in the $F814W$ and $F475W$ passbands are available at archive.stsci.edu/prepds/coma/datalist2.1.html. We have used data for the $F814W$ band, which guarantees the most comprehensive data for structural and luminosity function research (Carter *et al.*, 2008).

To ascertain which galaxies are members of our sample, we have used publicly available spectroscopic data from Mobasher *et al.* (2001), Michard, R. & Andreon, S. (2008), (Mahajan *et al.* (2011, 2010)) and Chiboucas *et al.* (2011). We have used available spectroscopic data that is available for galaxies brighter than F814W 19.5 (Mobasher 2001) and F814W 20.75 (den Brok 2014). We have used a limiting magnitude error of 0.01 and hence F814W 19.5 was set as the magnitude limit for our study. As dwarfs are defined on the basis of their absolute magnitudes, for which we require spectroscopy (redshift), our dwarf sample is 100% spectroscopically complete within the limiting magnitude of F814W 19.5.

Our sample was also matched with (den Brok *et al.* (2011, 2014)) where the Trenthem and Ferguson method was used to assign classes to members. In this method, class 0 is a spectroscopic member, class 1 is an almost certain member, class 2 is a likely member and class 3 is a probable member for galaxies $F814W < 20.75^m$. The redshift range of $0.023 \pm 0.009$ (Mobasher *et al.*, 2001), was chosen as the criterion for spectroscopic membership. In this study, we have used $19.5^m$ as the limiting magnitude as spectroscopic data is available only till this limit. We found 132 cluster members and using an absolute magnitude limit $F814W \geq -18.5^m$ (Marinova *et al.*, 2012) we have segregated dwarfs and non-dwarf galaxies. Our sample comprises of 78 dwarfs galaxies.

Further information regarding the data description, the SExtractor catalogue, and cluster membership are available in Poloji *et al.* (2022).

## 3. Structural decomposition and GALFIT results

For the 78 dwarf galaxies, we used GALFIT (Peng *et al.*, 2002) to perform a single Sérsic as well as a two-dimensional bulge disc decomposition. However, bulge disk decomposition could be done successfully only for 22% of objects. Therefore, in this paper, will consider single Sérsic fits for the entire sample of 78 dwarfs.

Sérsic profiles are characterised by,

$$\Sigma(r) = \Sigma_e e^{-\kappa[(r/r_e)^{1/n}-1]}$$

Where $r_e$ is the effective radius of galaxy where lies half of the total flux. $\Sigma_e$ is the pixel surface brightness at $r_e$ and $n$ is the Sérsic index. The relation between the effective radius $r_e$ and scale length $r_s$ are given by

$$r_e = 1.678 r_s.$$

The scale length $r_s$ is the radius at which intensity drops to $e^{-1}$.

For the input parameters of GALFIT, we used values from the SExtractor catalogues Hammer *et al.* (2010). The point-spread function (PSF) was generated by Tinytim (Krist *et al.*, 2011). Masking was done for objects in the frame that did not belong to the galaxy using the Image Reduction and Analysis Facility (IRAF).

Figure 2 shows a few examples of our GALFIT results: science image, model image and residual. The residuals are very useful to identify features of the galaxy and have been used to classify galaxies visually and are shown in the figure. Table 1 describes the column data from detailed analysis and the errors in fitting for our dwarf sample. The complete table is provided in the appendix (Table 2) .

[1]Hubble Space Telescope-Advanced Camera for Surveys

| Col | Parameter | description |
|---|---|---|
| 1 | COMA_ID | Name of Source. |
| 2 | RA (J2000) | Right ascension of Source |
| 3 | Dec (J2000) | declination of Source |
| 4 | Morphology | Based on visual inspection |
| 5 | $Gm$ | Galaxy magnitude |
| 6 | $Gm_{error}$ | Error in Galaxy magnitude |
| 7 | $G\mu_e$ | Galaxy surface brightness |
| 8 | $Gr_e$ | Galaxy effective radius in arcsec |
| 9 | $Gn$ | Galaxy Sérsic index |
| 10 | $Gn_{error}$ | Error in Galaxy Sérsic index |
| 11 | $Gb/a$ | Galaxy axis ratio |
| 12 | $Gb/a_{error}$ | Error in Galaxy axis ratio |
| 13 | $GPA$ | Galaxy position angle |
| 14 | $GPA_{error}$ | Error in Galaxy position angle |

**Table 1.** Results of single Sérsic fit for dwarfs.

We have classified our sample of dwarf galaxies based on visual inspection of the science, model residual images by the first author and confirmed by the second and third authors. (Fig.2). We have classified them as: ellipticals, lenticulars, spirals, barred spirals and irregulars. We found the following distribution: dwarf lenticular (*dS0*) 22%, dwarf Elliptical (*dE*) 69%, dwarf spirals (*dSp*) 4%, dwarf ring (*dring*) 1%, dwarf barred spirals (*dSBp*) 3% and dwarf irregular (*dIrr*) 1% galaxies. The distribution agrees with Michard, R. & Andreon, S. (2008) except only for 2 galaxies. Of these, one was classified as a spiral(due to the GALFIT residual). Michard, R. & Andreon, S. (2008) classified the dwarf galaxies in only two bins: Ellipticals (E0,...,E4) and Lenticulars.

## 4. Analysis and Discussion

### 4.1 *Colour Magnitude Relation (CMR)*

In this section we present the CMR for dwarf galaxies in the Coma Cluster with varied morphologies. The CMR and its scatter is a very good diagnostic tool as a star formation tracer (Visvanathan & Sandage, 1977; Bower *et al.*, 1992; Terlevich *et al.*, 2001; Hammer *et al.*, 2010; Poloji *et al.*, 2022; Hammer *et al.*, 2010; Poloji *et al.*, 2022). Colours ($F475W - F814W$) have been determined from the SExtractor catalogue and the extinction corrections have been made using values from Hammer *et al.* (2010). Figure 3 shows that except for one Irr galaxy, all the dwarf galaxies are red in colour, where $F475W - F814W \geq 0.811$. We observe that *dS0*, *dSp*, *dIrr* and *dring* galaxies are brighter than *dE* galaxies. The colour gets redder with magnitude and indicates that most of the *dS0*, *dSp* galaxies are redder than the *dE* galaxies. In the case of non-dwarf galaxies (Poloji *et al.*, 2022), spiral galaxies are blue while ellipticals are redder and brighter. The CMR indicates that there is no evidence of star forming activity in dwarf galaxies, including *dSp* galaxies as they have red colours between ≈ 0.8 to 1.3. Mahajan *et al.* (2011) suggest that star formation could have been quenched within the last Gyr in the central region of Coma Cluster. Galaxies may have lost their gas and stopped star formation activity due to ram-pressure striping (Boselli *et al.*, 2008). With this we infer that dwarf galaxies may have a reverse process of formation and evolution of non dwarf galaxies. These results are in agreement with Boselli *et al.* (2008) and Graham & Guzmán (2003). Dwarf galaxies in the central region of Coma Cluster show a smaller scatter compared to the galaxies in the Virgo cluster (Hamraz *et al.*, 2019) and these galaxies are in redder region which explains that these are older galaxies.

### 4.2 *Colour Distance Relation (from the centre NGC 4874)*

The relation between the colour of a galaxy and it's distance from the centre is an indicator of the formation and evolution of galaxies in the core and outskirts of a cluster. NGC 4874 is a giant elliptical galaxy, about ten times larger than the Milky Way, at the centre of the Coma Cluster and hence has a significant affect on the evolution of galaxies in the Coma Cluster.

Figure 4 shows the relation between the colour of a galaxy (F475W-F814W) and it's distance from NGC 4874

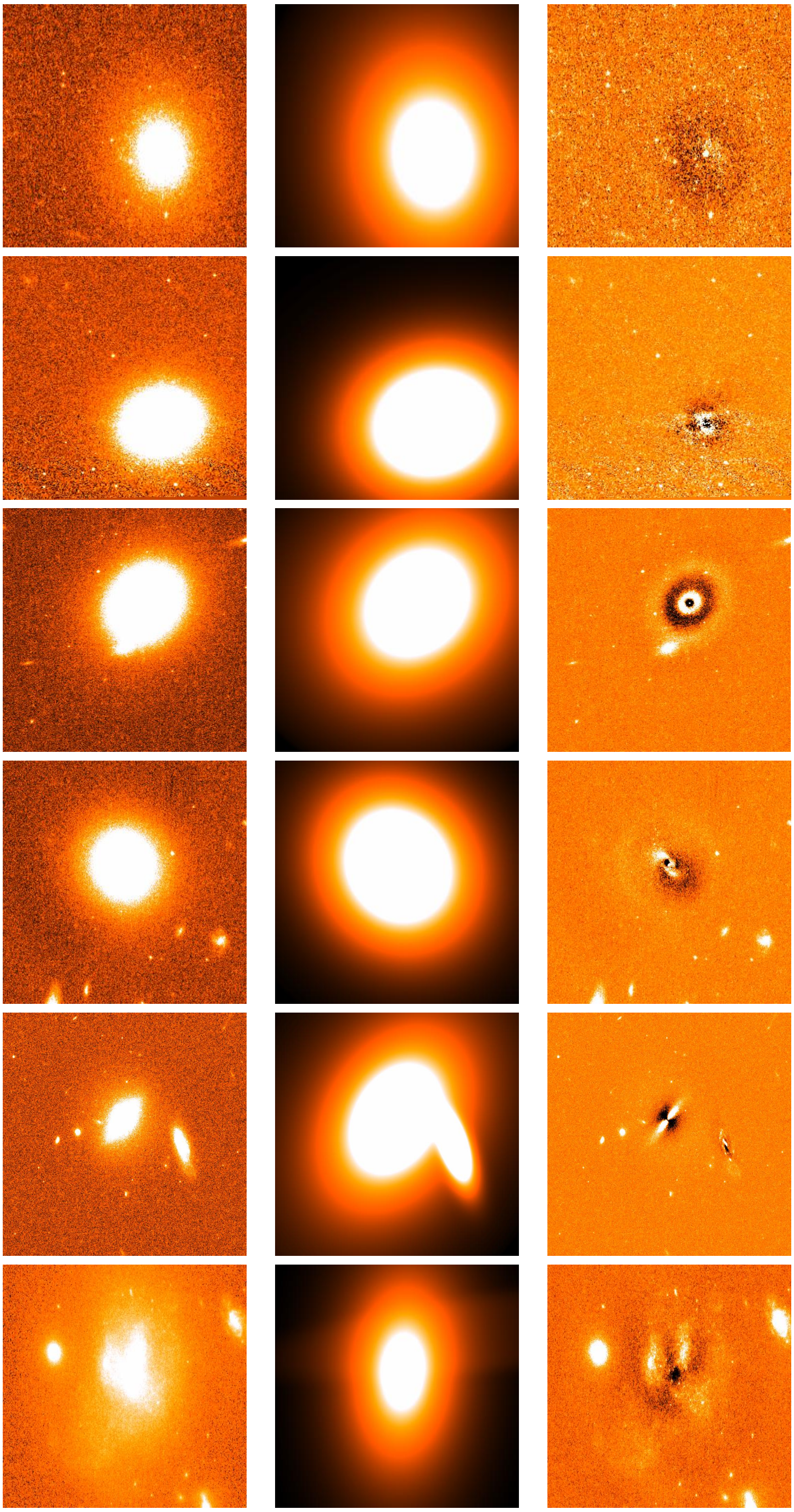

**Figure 2**. Sample results of GALFIT. The first column shows HST-ACS science images, the middle column shows GALFIT model images and the third column shows residuals which are obtained by subtracting the model image from the science image. The galaxy in the the first row has a zero residual and is classified as dwarf elliptical (*dE*), second row is a dwarf lenticular (*dS0*), third row is a dwarf ring (*dring*), fourth row is dwarf spiral (*dSp*), fifth is dwarf barred spiral (*dSBp*) and last is dwarf irregular (*dIrr*).

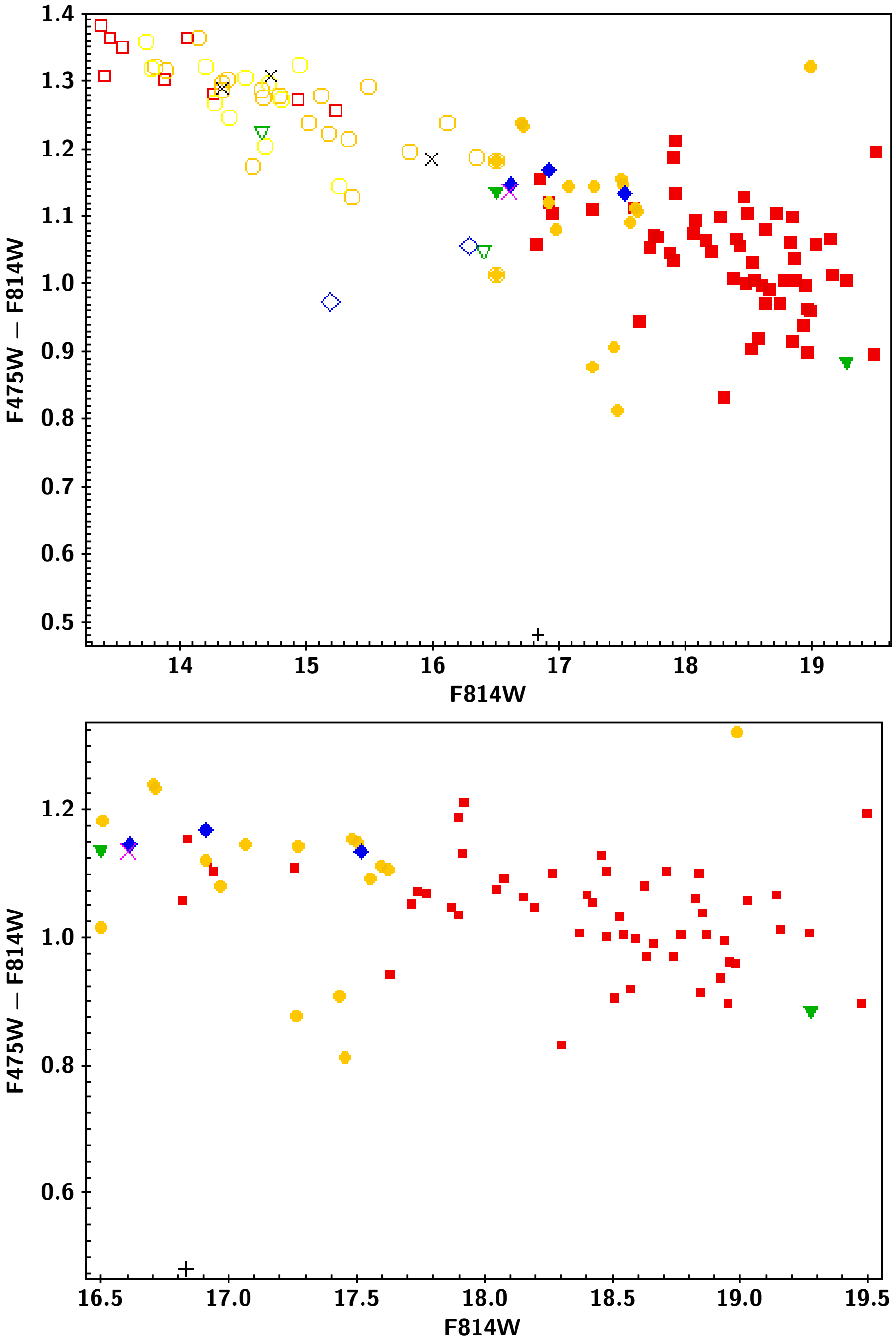

**Figure 3**. Colour Magnitude Relation for dwarf member galaxies. Here the red filled squares represents the *dE* galaxies, yellow filled circles are *dS0*, pink cross mark is *dring*, blue filed rhombus are *dSp*, green triangles are *dSBp*, black plus mark is *dIrr*, black cross marks are *ring* and open symbols in the same colours are the non-dwarf galaxies of same morphological type. The lower plot is a zoomed version for dwarfs only.

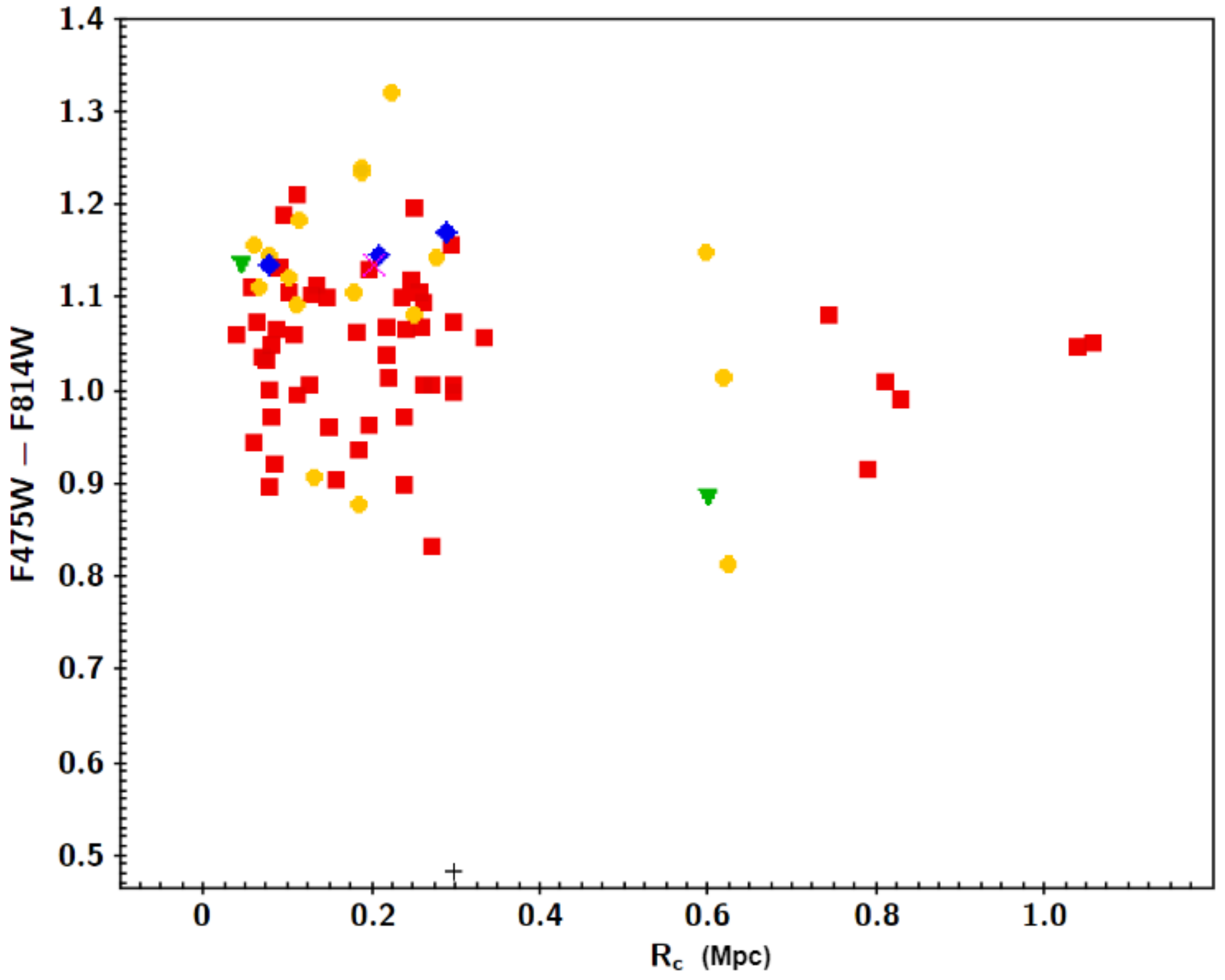

**Figure 4**. The colour versus distance from the centre ($R_c$ in Mpc) (NGC 4874) relation for dwarf member galaxies. Here The red filled squares represents the *dE* galaxies, yellow filled circles are *dS0*, pink cross mark is *dring*, blue filed rhombus are *dSp*, green triangles are *dSBp*, black plus mark is *dIrr*.

($R_c$). Due to the lack of data in the complete cluster region, where 19 tiles were within 0.5 Mpc and the remaining 6 tiles were from 0.9 to 1.75 Mpc, we do not have good coverage in Rc in the Figure indicated by the empty regions in Fig. 1. The empty regions were not covered by the survey. In this figure, we observe that most of the dwarf galaxies are within 0.33 Mpc (except for one *dIrr*) and are placed in the redder region. The three *dSp* and one *dSBp* galaxies are within 0.33 Mpc and the colour is $F475W - F814W \geq 1.13$ mag. Generally, spiral galaxies show star formation activity e.g. non-dwarf spirals in the core of Coma cluster are in the blue region (Poloji *et al.*, 2022). *dSp*s in Coma seem to have different properties than their non-dwarf counterparts. *dE* and *dS0* galaxies do not show any difference in colour (Fig. 4), they are equally distributed in colour $F475W - F814W \geq 0.8$ mag.

### 4.3 *Kormendy relation*

Kormendy relation (Kormendy, 1977, 1985) is a plot between surface brightness and effective radius. It is used to study the properties of elliptical galaxies. From the Kormendy relation (Fig. 5), we observe that as effective radius increases surface brightness decreases. We also observe that *dE* (red filled circles) galaxies have low surface brightness compared to other dwarfs. If we compare with non-dwarf galaxies, the elliptical (red open circles) galaxies a have a higher surface brightness and higher effective radius than the other galaxies (Poloji *et al.*, 2022). The Kormendy relation shows that spirals, barred spirals and faint lenticulars occupy similar regions both for dwarfs and non-dwarfs. This suggests that dwarf galaxies may have formed from non-dwarfs in the case of late type galaxies.

George (2017) studied a sample of 55 star-forming elliptical galaxies. He found the 32 of them showed a bulge Sérsic Index $n > 2$ and are bulge-dominated. We have studied the core of Coma Cluster and found most of the dwarfs are ellipticals, and red indicative of low star formation activity. As bulge-disk decomposition was not possible for our sample, we do not have information of our bulge Sérsic Indices.

Mahajan *et al.* (2018) studied the Coma Supercluster which comprises the two clusters, Coma (Abell 1656) and Abell 1367, and found that dwarfs are almost always blue except in the densely populated clusters and groups. Our region of study, is close to the core, hence denser, and agrees with their findings as we find the dwarf population to be red.

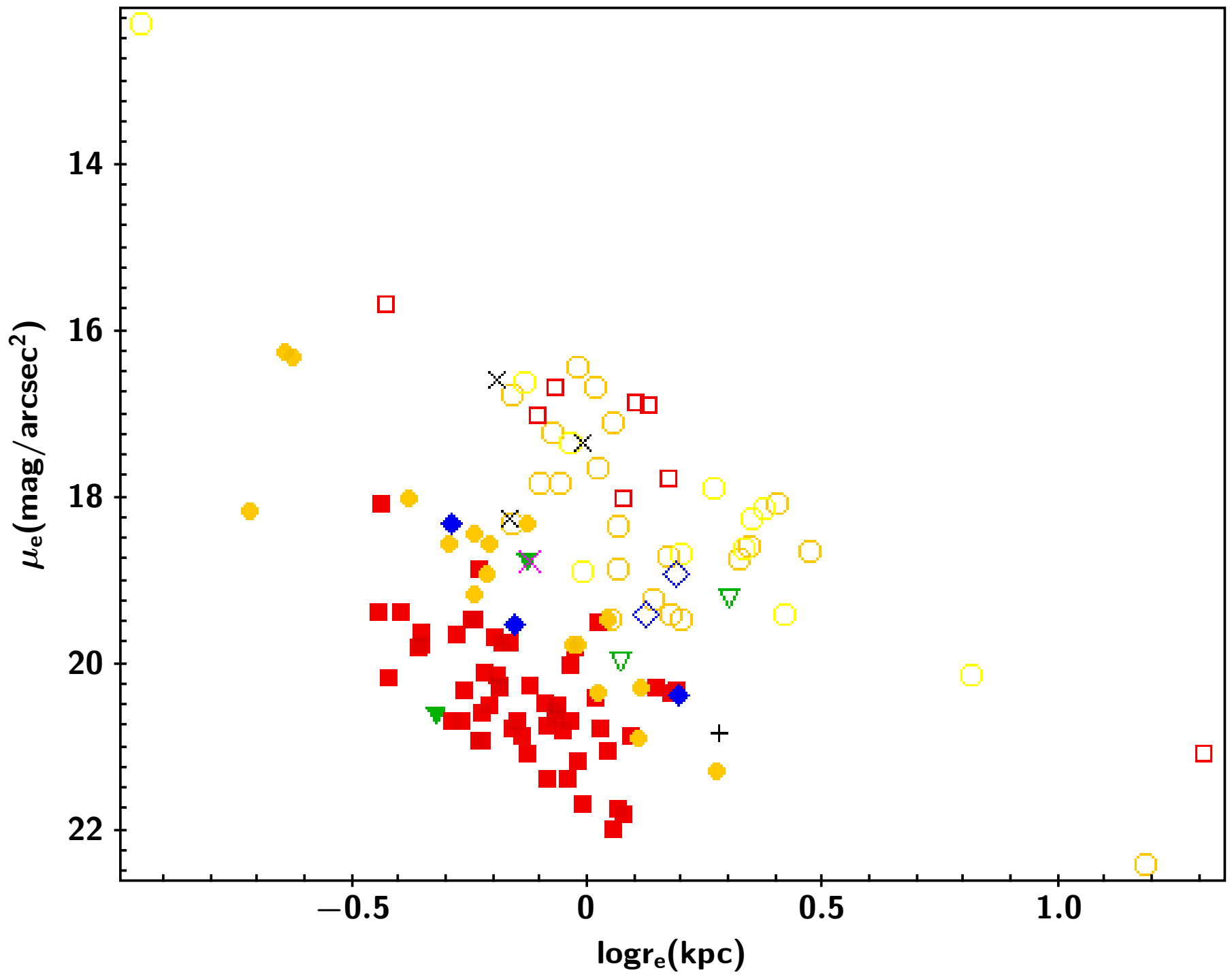

**Figure 5**. The Kormendy relation for dwarf member galaxies. Here the red filled squares represents the *dE* galaxies, yellow filled circles are *dS0*, pink cross mark is *dring*, blue filed rhombus are *dSp*, green triangles are *dSBp*, black plus mark is *dIrr*, black cross marks are *ring* and open symbols in the same colours are the non-dwarf galaxies of same morphological type.

## 5. Conclusions

In this work, we studied the morphological types and properties of dwarf galaxies in the core of Coma Cluster using HST/ACS data. We obtain the structural properties using GALFIT in a single Sérsic fit. We have classified galaxies based on visual inspection of residuals of GALFIT results. We find the following distribution: dwarf spirals (*dSp*) 4%, dwarf lenticular (*dS0*) 22%, dwarf Elliptical (*dE*) 69%, dwarf ring (*dring*) 1%, dwarf barred spirals (*dSBp*) 3% and dwarf irregular (*dIrr*) 1% galaxies.

Using the CMR for dwarf galaxies, we observe that all the galaxies (except one *dIrr* galaxy) are in the redder region of the colour-Magnitude Diagram which suggests that these galaxies are old and have low star formation activity (Hamraz *et al.*, 2019). We also observe that *dE* galaxies are fainter than the *dSp*, *dS0* and *dSBp* galaxies. Generally, ellipticals are brighter and redder compared to other galaxies (Poloji *et al.*, 2022). Smith *et al.* (2008) found that the age of red-sequence galaxies depends on their location within clusters, specifically for Coma cluster. They also concluded that dwarf galaxies are especially susceptible to environmental 'quenching' and the south-western part of Coma is an example of recent quenching in an infalling subcluster. Figure 3 shows that *dE*s are bluer than the *dSp*, *dSBp* and *dS0* galaxies, which could be a possible indicator of their star formation activity.

From the relation between colour of the galaxy and distance from the giant elliptical galaxy NGC 4874, in the central region of the cluster, we observe that the galaxies in our sample are red in colour, irrespective of morphological type and distance from centre. Also, for dwarfs, there is no clear distinction between morphological type and distance from the centre. In the case of non-dwarfs, we found that the central region was populated by early type galaxies while the outskirts had late type galaxies Poloji *et al.* (2022).

From the Kormendy relation, we observe that the *dE* galaxies have a larger surface brightness and smaller effective radius while *dSp* and *dSBp* galaxies have smaller surface brightness and larger effective radius. The *dSp*, *dSBp* galaxies show similar properties as late type spirals and faint lenticulars (Poloji *et al.*, 2022). This reverse trend in dwarf galaxies compared to non dwarf galaxies was also found by Boselli *et al.* (2008); Graham & Guzmán (2003). There has been a debate on whether *dE* galaxies have properties similar to low luminosity giant elliptical galaxies or *dIrr* galaxies (Binggeli *et al.*, 1985; Kormendy, 1985; Binggeli & Cameron, 1991; Jerjen & Binggeli, 1997; Aguerri *et al.*, 2005). Detailed studies of cluster dwarfs are required to shed more light on these investigations.

## Acknowledgements

The authors gratefully thank the referee for the constructive comments and recommendations which definitely help to improve the readability and quality of the paper.

## Data Availability

Table 2 shows the detailed results obtained in the paper. The data used had been added to the Appendix and the table has been described in Table 1. Soft copies can be obtained on request.

## References

Aguerri, J. A. L., Iglesias-Páramo, J., Vílchez, J. M., Muñoz-Tuñón, C., & Sánchez-Janssen, R. 2005, AJ, 130, 475

Andreon, S. 1996, A&A, 314, 763

Bertin, E., & Arnouts, S. 1996, A&AS, 117, 393

Binggeli, B., & Cameron, L. M. 1991, A&A, 252, 27

Binggeli, B., Sandage, A., & Tammann, G. A. 1985, AJ, 90, 1681

Boselli, A., Boissier, S., Cortese, L., & Gavazzi, G. 2008, A&A, 489, 1015

Bower, R. G., Lucey, J. R., & Ellis, R. S. 1992, MNRAS, 254, 589

Carter, D., Goudfrooij, P., Mobasher, B., *et al.* 2008, ApJS, 176, 424

Chiboucas, K., Tully, R. B., Marzke, R. O., *et al.* 2011, ApJ, 737, 86

Dekel, A., & Silk, J. 1986, ApJ, 303, 39

den Brok, M., Peletier, R. F., Valentijn, E. A., *et al.* 2011, MNRAS, 414, 3052

den Brok, M., Peletier, R. F., Seth, A., *et al.* 2014, MNRAS, 445, 2385

George, K. 2017, A&A, 598, A45

Graham, A. W., & Guzmán, R. 2003, AJ, 125, 2936

Gutiérrez, C. M., Trujillo, I., Aguerri, J. A. L., Graham, A. W., & Caon, N. 2004, ApJ, 602, 664

Hammer, D., Verdoes Kleijn, G., Hoyos, C., *et al.* 2010, ApJS, 191, 143

Hamraz, E., Peletier, R. F., Khosroshahi, H. G., *et al.* 2019, A&A, 625, A94

Hasan, S. N., Shah, P., & Nagamani, P. 2019, in Dwarf Galaxies: From the Deep Universe to the Present, ed. K. B. W. McQuinn & S. Stierwalt, Vol. 344, 417–419

Jerjen, H., & Binggeli, B. 1997, in Astronomical Society of the Pacific Conference Series, Vol. 116, The Nature of Elliptical Galaxies; 2nd Stromlo Symposium, ed. M. Arnaboldi, G. S. Da Costa, & P. Saha, 239

Komiyama, Y., Sekiguchi, M., Kashikawa, N., *et al.* 2002, ApJS, 138, 265

Kormendy, J. 1977, ApJ, 218, 333

—. 1985, ApJ, 295, 73

Kourkchi, E., Khosroshahi, H. G., Carter, D., & Mobasher, B. 2012, MNRAS, 420, 2835

Krist, J. E., Hook, R. N., & Stoehr, F. 2011, in Society of Photo-Optical Instrumentation Engineers (SPIE) Conference Series, Vol. 8127, Optical Modeling and Performance Predictions V, ed. M. A. Kahan, 81270J

Mahajan, S., Haines, C. P., & Raychaudhury, S. 2010, VizieR Online Data Catalog, J/MNRAS/404/1745

—. 2011, MNRAS, 412, 1098

Mahajan, S., Singh, A., & Shobhana, D. 2018, Monthly Notices of the Royal Astronomical Society, 478, 4336

Marinova, I., Jogee, S., Weinzirl, T., *et al.* 2012, ApJ, 746, 136

Michard, R., & Andreon, S. 2008, A&A, 490, 923

Mobasher, B., Bridges, T. J., Carter, D., *et al.* 2001, ApJS, 137, 279

Peng, C. Y., Ho, L. C., Impey, C. D., & Rix, H.-W. 2002, AJ, 124, 266

Poloji, N., Hasan, P., & Hasan, S. 2022, New Astronomy, 101963

Poloji, N., Hasan, P., & Hasan, S. N. 2022, MNRAS, 510, 4463

Poloji, N., Hasan, P., & Hasan, S. N. 2023, Research in Astronomy and Astrophysics, 23, 065019

Secker, J., Harris, W. E., & Plummer, J. D. 1997, Publications of the Astronomical Society of the Pacific, 109, 1377

Smith, R. J., Marzke, R. O., Hornschemeier, A. E., *et al.* 2008, MNRAS, 386, L96

Terlevich, A. I., Caldwell, N., & Bower, R. G. 2001, MNRAS, 326, 1547

Thompson, L. A., & Gregory, S. A. 1993, AJ, 106, 2197

Visvanathan, N., & Sandage, A. 1977, ApJ, 216, 214

| | Position | | | Galaxy properties | | | | | | | | | |
|---|---|---|---|---|---|---|---|---|---|---|---|---|---|
| **COMAID** | **RA** | **Dec** | **Morphology** | $Gm$ | $Gm_{err}$ | $G\mu_e$ | $Gr_e$ | $Gn$ | $Gn_{err}$ | $Gb/a$ | $Gb/a_{err}$ | $G_{PA}$ | $G_{PAerr}$ |
| **COMAi125927.694p28145.91** | **194.8653924** | **28.029421** | *dE* | **18.86040039** | **0.02** | **19.38575669** | **0.735378** | **1.1164** | **0.05** | **0.9107** | **0.02** | **37.4584** | **7.57** |
| **COMAi13007.123p275551.49** | **195.0296795** | **27.9309714** | *dE* | **17.49999924** | **0.0** | **18.05940538** | **0.747** | **2.7033** | **0.03** | **0.9684** | **0.0** | **12.06** | **3.91** |
| **COMAi13032.520p28201.50** | **195.1355009** | **28.0337524** | *dE* | **19.51259918** | **0.01** | **20.16144435** | **0.77841** | **2.5527** | **0.1** | **0.966** | **0.02** | **-86.127** | **16.75** |
| **COMAi125949.960p275433.04** | **194.9581696** | **27.9091786** | *dE* | **18.59999962** | **0.01** | **19.37507** | **0.825** | **0.9** | **0.01** | **0.9951** | **0.0** | **68.32** | **0.1** |
| **COMAi13013.809p28243.59** | **195.0575408** | **28.0454418** | *dE* | **18.83870049** | **0.1** | **19.81132706** | **0.903576** | **1.4798** | **0.04** | **0.9838** | **0.01** | **-21.0896** | **20.21** |
| **COMAi13004.034p28030.79** | **195.0168123** | **28.008554** | *dE* | **18.77430077** | **0.01** | **19.77232186** | **0.914205** | **1.315** | **0.03** | **0.9457** | **0.01** | **43.3991** | **6.43** |
| **COMAi125902.433p28021.34** | **194.7601385** | **28.005928** | *dE* | **18.59419937** | **0.01** | **19.60042098** | **0.917664** | **1.0217** | **0.03** | **0.7574** | **0.01** | **-0.4042** | **1.51** |
| **COMAi13037.300p275441.05** | **195.1554192** | **27.9114034** | *dE* | **19.37140007** | **0.01** | **20.68095532** | **1.055238** | **1.0283** | **0.03** | **0.9443** | **0.01** | **85.5239** | **6.69** |
| **COMAi13020.398p28414.03** | **195.0849926** | **28.0705661** | *dE* | **18.2618** | **0.02** | **19.63383025** | **1.086039** | **1.0428** | **0.04** | **0.9323** | **0.01** | **-55.3851** | **6.68** |
| **COMAi13027.880p275916.52** | **195.1161674** | **27.9879235** | *dE* | **19.23660011** | **0.01** | **20.66686013** | **1.115556** | **1.1144** | **0.03** | **0.7872** | **0.01** | **38.8477** | **1.83** |
| **COMAi13042.519p28325.39** | **195.1771641** | **28.057054** | *dE* | **18.88429947** | **0.01** | **20.32377235** | **1.120299** | **1.5843** | **0.04** | **0.8852** | **0.01** | **5.2075** | **2.06** |
| **COMAi125955.701p275503.77** | **194.9820889** | **27.9177163** | *dE* | **17.94410057** | **0.01** | **19.46586815** | **1.163571** | **1.1457** | **0.02** | **0.9492** | **0.01** | **-35.642** | **4.27** |
| **COMAi125937.208p275213.73** | **194.9050355** | **27.8704823** | *dE* | **17.90229912** | **0.01** | **19.45570811** | **1.18065** | **1.41** | **0.01** | **0.9** | **0.0** | **-39.266** | **2.55** |
| **COMAi125955.941p275748.79** | **194.9830877** | **27.9635535** | *dE* | **19.31910057** | **0.01** | **20.91609998** | **1.20459** | **1.0052** | **0.03** | **0.9048** | **0.01** | **-31.8808** | **3.56** |
| **COMAi125928.502p28109.38** | **194.8687621** | **28.0192743** | *dE* | **17.22930069** | **0.01** | **18.84651275** | **1.215855** | **1.1663** | **0.02** | **0.9409** | **0.01** | **23.4474** | **5.23** |
| **COMAi13018.715p275512.63** | **195.0779805** | **27.9201752** | *dE* | **18.93970032** | **0.01** | **20.57553136** | **1.226325** | **1.7124** | **0.04** | **0.8719** | **0.01** | **9.1381** | **2.56** |
| **COMAi125943.535p275620.67** | **194.9313991** | **27.9390767** | *dE* | **19.27830048** | **0.01** | **20.91765062** | **1.228314** | **0.7153** | **0.02** | **0.9896** | **0.01** | **-20.568** | **37.42** |
| **COMAi125713.240p272437.23** | **194.3051686** | **27.4103435** | *dE* | **18.44250031** | **0.01** | **20.09437834** | **1.235421** | **1.1658** | **0.02** | **0.8619** | **0.01** | **-42.4102** | **1.79** |
| **COMAi125711.022p273142.31** | **194.2959255** | **27.5284218** | *dE* | **18.76909943** | **0.01** | **20.48356** | **1.271544** | **1.1927** | **0.03** | **0.726** | **0.01** | **12.702** | **0.99** |
| **COMAi125933.238p28152.55** | **194.8884923** | **28.031265** | *dE* | **17.89610024** | **0.01** | **19.66208838** | **1.302078** | **1.7672** | **0.02** | **0.9319** | **0.01** | **-53.0179** | **4.99** |
| **COMAi13048.042p28557.40** | **195.200177** | **28.0992783** | *dE* | **18.33209915** | **0.01** | **20.11880601** | **1.314561** | **1.4905** | **0.03** | **0.9786** | **0.01** | **-22.9018** | **11.6** |
| **COMAi13014.170p28407.28** | **195.0590423** | **28.0686909** | *dE* | **18.42800064** | **0.02** | **20.24295318** | **1.331772** | **0.7328** | **0.03** | **0.7253** | **0.01** | **-53.4227** | **1.45** |
| **COMAi13036.582p275552.24** | **195.1524263** | **27.9311799** | *dE* | **18.4364006** | **0.01** | **20.27090924** | **1.34382** | **1.8022** | **0.03** | **0.7571** | **0.01** | **-69.0593** | **1.04** |
| **COMAi13000.989p275929.75** | **195.0041249** | **27.9915986** | *dE* | **17.88300056** | **0.01** | **19.73412154** | **1.35414** | **1.556** | **0.02** | **0.9227** | **0.0** | **-66.3066** | **2.26** |
| **COMAi13026.155p28032.00** | **195.1089811** | **28.0088916** | *dE* | **17.79930038** | **0.01** | **19.74355647** | **1.413483** | **1.366** | **0.02** | **0.8304** | **0.0** | **-2.0206** | **0.89** |
| **COMAi13005.347p275628.95** | **195.0222803** | **27.9413765** | *dE* | **18.80000038** | **0.01** | **20.76623013** | **1.427859** | **1.12** | **0.03** | **0.9042** | **0.01** | **69.3042** | **1.26** |
| **COMAi125844.379p274740.91** | **194.6849141** | **27.7946973** | *dE* | **18.65840073** | **0.01** | **20.66406771** | **1.454028** | **1.1748** | **0.03** | **0.5548** | **0.0** | **-89.1381** | **0.49** |
| **COMAi125942.880p28202.20** | **194.9286688** | **28.0339468** | *dE* | **18.80040092** | **0.01** | **20.85352** | **1.486188** | **0.7615** | **0.02** | **0.9582** | **0.01** | **56.5261** | **10.23** |
| **COMAi125905.940p28228.83** | **194.7747534** | **28.041344** | *dE* | **18.9432003** | **0.01** | **21.06116304** | **1.5312** | **0.6995** | **0.03** | **0.8826** | **0.01** | **19.8424** | **3.51** |
| **COMAi13039.113p28035.46** | **195.162974** | **28.0098516** | *dE* | **18.11570091** | **0.01** | **20.24823296** | **1.541508** | **1.2451** | **0.02** | **0.8787** | **0.01** | **4.2093** | **2.27** |
| **COMAi125921.612p28102.18** | **194.8400509** | **28.0172748** | *dE* | **18.12970085** | **0.01** | **20.4485191** | **1.679589** | **0.8719** | **0.3** | **0.8524** | **0.01** | **-81.0493** | **2.43** |
| **COMAi125958.218p275410.84** | **194.992576** | **27.9030126** | *dE* | **19.05020065** | **0.01** | **21.37178257** | **1.681728** | **0.9899** | **0.03** | **0.6331** | **0.01** | **30.6303** | **0.77** |
| **COMAi13011.407p275436.39** | **195.047533** | **27.9101096** | *dE* | **18.40610046** | **0.01** | **20.75026028** | **1.699305** | **1.9657** | **0.04** | **0.8006** | **0.01** | **-81.6598** | **0.99** |
| **COMAi13033.329p275849.31** | **195.1388745** | **27.9803641** | *dE* | **18.12580032** | **0.01** | **20.54275639** | **1.757238** | **1.3132** | **0.02** | **0.9497** | **0.01** | **-32.9603** | **4.69** |
| **COMAi13025.984p28344.73** | **195.1082682** | **28.0624277** | *dE* | **18.21339912** | **0.01** | **20.63937457** | **1.764552** | **1.2507** | **0.02** | **0.9217** | **0.01** | **-38.7117** | **3.5** |
| **COMAi13036.670p275427.51** | **195.1527932** | **27.9076443** | *dE* | **18.05860062** | **0.01** | **20.49884597** | **1.776186** | **1.7199** | **0.03** | **0.7931** | **0.0** | **-3.5794** | **0.81** |
| **COMAi13022.656p275754.88** | **195.0944002** | **27.9652463** | *dE* | **18.2807991** | **0.02** | **20.78282619** | **1.827447** | **1.8117** | **0.05** | **0.9187** | **0.01** | **78.4107** | **2.36** |
| **COMAi125828.358p271315.15** | **194.6181604** | **27.220876** | *dE* | **18.82549973** | **0.01** | **21.38505265** | **1.876506** | **1.0424** | **0.02** | **0.682** | **0.01** | **86.775** | **0.61** |
| **COMAi125946.712p28000.39** | **194.944636** | **28.0001106** | *dE* | **17.44570084** | **0.01** | **20.01608** | **1.885881** | **1.9453** | **0.01** | **0.8884** | **0.0** | **83.6071** | **1.56** |
| **COMAi125942.377p28158.58** | **194.9265713** | **28.0329406** | *dE* | **18.09179993** | **0.01** | **20.67132981** | **1.893849** | **1.3682** | **0.03** | **0.9031** | **0.01** | **14.3428** | **3.46** |
| **COMAi125946.943p275930.90** | **194.9455974** | **27.9919171** | *dE* | **17.16080017** | **0.01** | **19.78235049** | **1.930854** | **1.4** | **0.01** | **0.8343** | **0.0** | **35.1575** | **0.36** |
| **COMAi13032.603p28331.50** | **195.1358493** | **28.0587508** | *dE* | **18.50520058** | **0.01** | **21.15908353** | **1.959819** | **0.8561** | **0.02** | **0.9325** | **0.01** | **21.1011** | **4.68** |
| **COMAi125712.266p272313.37** | **194.3011109** | **27.3870485** | *dE* | **18.95110054** | **0.01** | **21.66122** | **2.011239** | **0.8354** | **0.02** | **0.7189** | **0.01** | **-76.363** | **0.9** |
| **COMAi125948.589p275858.05** | **194.9524563** | **27.9827944** | *dE* | **17.56689949** | **0.01** | **20.41683911** | **2.145** | **1.2** | **0.02** | **0.7912** | **0.0** | **89.301** | **1.27** |
| **COMAi13000.949p275643.85** | **195.0039561** | **27.9455141** | *dE* | **16.62239952** | **0.01** | **19.48475538** | **2.1573** | **2.8846** | **0.03** | **0.9018** | **0.0** | **82.26** | **1.05** |
| **COMAi125635.495p271430.39** | **194.1478996** | **27.2417767** | *dE* | **17.88230057** | **0.01** | **20.77085676** | **2.183487** | **1.0498** | **0.01** | **0.6131** | **0.0** | **83.1223** | **0.44** |
| **COMAi13042.881p28313.82** | **195.1786734** | **28.053839** | *dE* | **18.05239983** | **0.01** | **21.03568564** | **2.280849** | **1.2097** | **0.02** | **0.5713** | **0.0** | **38.1865** | **0.48** |
| **COMAi13032.493p275833.33** | **195.135389** | **27.9759271** | *dE* | **18.94370003** | **0.01** | **21.97319144** | **2.329902** | **0.9439** | **0.02** | **0.6483** | **0.01** | **-67.1164** | **0.72** |
| **COMAi13027.608p28323.85** | **195.1150344** | **28.056625** | *dE* | **18.64499969** | **0.01** | **21.73074703** | **2.391051** | **0.9726** | **0.02** | **0.6948** | **0.01** | **-28.2728** | **0.9** |
| **COMAi13044.130p28215.42** | **195.1838764** | **28.0376178** | *dE* | **18.66019936** | **0.01** | **21.80678277** | **2.458986** | **1.0588** | **0.03** | **0.7411** | **0.01** | **-22.0414** | **1.25** |

| | Position | | | Galaxy properties | | | | | | | | | |
|---|---|---|---|---|---|---|---|---|---|---|---|---|---|
| **COMAID** | **RA** | **Dec** | **Morphology** | $Gm$ | $Gm_{err}$ | $G\mu_e$ | $Gr_e$ | $Gn$ | $Gn_{err}$ | $Gb/a$ | $Gb/a_{err}$ | $G_{PA}$ | $G_{PAerr}$ |
| COMAi125636.785p271247.90 | 194.1532741 | 27.2133073 | *dE* | 17.64149971 | 0.01 | 20.85654 | 2.537739 | 1.1549 | 0.02 | 0.9597 | 0.01 | 50.4072 | 5.2 |
| COMAi125845.533p274513.75 | 194.6897239 | 27.7538197 | *dE* | 16.79869957 | 0.01 | 20.27290841 | 2.859441 | 2.1407 | 0.02 | 0.9532 | 0.0 | 23.6253 | 1.42 |
| COMAi125959.476p275626.02 | 194.99782 | 27.9405637 | *dE* | 16.67699928 | 0.01 | 20.32801139 | 3.102 | 2.114 | 0.03 | 0.8901 | 0.0 | -36.2141 | 0.69 |
| COMAi13034.430p275604.95 | 195.1434613 | 27.9347093 | *dE* | 16.59999962 | 0.0 | 20.30432165 | 3.179097 | 2.29 | 0.02 | 0.9028 | 0.0 | -2.0455 | 1.4 |
| COMAi125911.543p28033.32 | 194.7980988 | 28.0092582 | *dS0* | 16.18929977 | 0.0 | 18.30512999 | 1.529697 | 1.9175 | 0.01 | 0.9129 | 0.0 | 68.6911 | 0.8 |
| COMAi125820.530p272545.99 | 194.5855418 | 27.4294427 | *dS0* | 16.49450035 | 0.0 | 19.46910617 | 2.27175 | 1.6273 | 0.01 | 0.4812 | 0.0 | 64.2921 | 0.07 |
| COMAi125814.968p272744.84 | 194.5623698 | 27.4624564 | *dSBp* | 19.43220062 | 0.01 | 20.59820597 | 0.987735 | 0.9597 | 0.02 | 0.4082 | 0.0 | 21.8133 | 0.47 |
| COMAi125937.990p28003.52 | 194.9082917 | 28.0009788 | *dSBp* | 16.62399979 | 0.0 | 18.73283459 | 1.524777 | 1.5722 | 0.01 | 0.844 | 0.0 | 59.9722 | 0.5 |
| COMAi13027.339p28033.40 | 195.1139143 | 28.0092797 | *dS0* | 19.00890083 | 0.01 | 18.17414738 | 0.393087 | 1.9631 | 0.04 | 0.9082 | 0.01 | -21.0309 | 4.41 |
| COMAi13018.883p28033.55 | 195.07868 | 28.0093198 | *dS0* | 16.71049995 | 0.0 | 16.25714768 | 0.468564 | 2.9467 | 0.02 | 0.8667 | 0.0 | -63.7687 | 0.91 |
| COMAi13018.873p28033.38 | 195.0786388 | 28.0092725 | *dS0* | 16.67509956 | 0.0 | 16.31419386 | 0.488943 | 3.1844 | 0.02 | 0.8835 | 0.0 | -64.0868 | 1.22 |
| COMAi125937.010p28106.95 | 194.904209 | 28.0185978 | *dS0* | 17.14989967 | 0.01 | 18.00395009 | 0.855558 | 1.9981 | 0.03 | 0.9315 | 0.01 | 8.4757 | 2.68 |
| COMAi13017.641p275915.27 | 195.0735082 | 27.9875773 | *dS0* | 17.24980087 | 0.0 | 18.54482479 | 1.0482 | 2.52 | 0.04 | 0.93 | 0.01 | -39.2255 | 2.0 |
| COMAi125930.270p28115.17 | 194.8761261 | 28.020883 | *dS0* | 16.86819954 | 0.01 | 18.42075865 | 1.180188 | 1.5818 | 0.02 | 0.8077 | 0.01 | 87.3221 | 1.35 |
| COMAi125815.292p272753.05 | 194.5637167 | 27.4647363 | *dS0* | 17.5939991 | 0.01 | 19.15419522 | 1.184346 | 2.0891 | 0.02 | 0.7422 | 0.0 | -42.3585 | 0.75 |
| COMAi125926.458p275124.81 | 194.8602453 | 27.8568931 | *dS0* | 17.21580048 | 0.01 | 18.91458854 | 1.2624 | 1.9 | 0.02 | 0.9894 | 0.0 | 68.94 | 3.2 |
| COMAi125931.893p275140.76 | 194.8828881 | 27.8613245 | *dS0* | 16.83870049 | 0.0 | 18.55598677 | 1.2732 | 1.9362 | 0.01 | 0.9119 | 0.0 | -9.44 | 0.9 |
| COMAi13030.954p28630.22 | 195.1289774 | 28.1083964 | *dS0* | 17.15079994 | 0.0 | 19.76546682 | 1.924743 | 2.0091 | 0.02 | 0.5402 | 0.0 | -25.4477 | 0.19 |
| COMAi125950.183p275445.52 | 194.959098 | 27.9126465 | *dS0* | 17.09579964 | 0.0 | 19.75902283 | 1.968267 | 1.2 | 0.01 | 0.8139 | 0.0 | 33.6638 | 0.3 |
| COMAi13005.685p275535.19 | 195.0236882 | 27.9264442 | *dS0* | 17.46000023 | 0.0 | 20.33049403 | 2.1654 | 1.06 | 0.01 | 0.5718 | 0.0 | 58.507 | 0.17 |
| COMAi125831.666p272342.04 | 194.6319424 | 27.3950124 | *dS0* | 17.57919998 | 0.01 | 20.88177403 | 2.64213 | 0.928 | 0.01 | 0.4481 | 0.0 | -89.376 | 0.22 |
| COMAi13035.420p275634.06 | 195.1475863 | 27.942797 | *dS0* | 16.95869942 | 0.0 | 20.27192578 | 2.655123 | 1.85 | 0.01 | 0.539 | 0.0 | 12.653 | 0.22 |
| COMAi13013.395p28311.82 | 195.0558151 | 28.0532851 | *dS0* | 17.14099998 | 0.0 | 21.27891296 | 3.8817 | 1.8671 | 0.01 | 0.2992 | 0.0 | -21.5725 | 0.09 |
| COMAi13041.193p28242.34 | 195.1716389 | 28.0450969 | *dSp* | 17.00590057 | 0.01 | 18.31348561 | 1.054281 | 1.1247 | 0.01 | 0.8704 | 0.0 | 64.0527 | 1.22 |
| COMAi125953.929p275813.75 | 194.974706 | 27.9704888 | *dSp* | 17.53340073 | 0.01 | 19.52157823 | 1.442364 | 1.1446 | 0.01 | 0.9133 | 0.0 | -65.7515 | 0.74 |
| COMAi13024.823p275535.94 | 195.1034295 | 27.9266523 | *dSp* | 16.65149994 | 0.0 | 20.37282864 | 3.204093 | 1.5898 | 0.01 | 0.3803 | 0.0 | 58.9809 | 0.11 |
| COMAi125842.638p274537.86 | 194.6776601 | 27.7605173 | *dIrr* | 16.68040009 | 0.01 | 20.82519467 | 3.894021 | 1.268 | 0.01 | 0.787 | 0.0 | -12.6901 | 0.31 |
| COMAi13021.673p275354.81 | 195.0903082 | 27.8985594 | *dring* | 16.63400002 | 0.0 | 18.77711351 | 1.549038 | 1.7727 | 0.01 | 0.7459 | 0.0 | -17.9919 | 0.29 |

**Table 2: Results of single Sérsic structural decomposition of 78 dwarf member galaxies. The table description can be seen in Table 1**